\documentstyle[12pt,titlepage,epsf,amssymb]{article}
\title{IMPROVED CONSTRAINTS ON $\bar B^{0} \rightarrow \rho^{+} \ell^{-} \bar
\nu_{\ell}$ FORM FACTORS}
\author{Damir Becirevic}
\date{July, 1996}
\def\br{B \rightarrow \rho \ell \bar\nu_{\ell} }
\begin{document}
\begin{titlepage}
\begin{flushright}
Preprint LPTHE - Orsay 96/52 \\
hep-ph/9607243
\end{flushright}
\begin{center}
\vspace{1in}
  {\bf {\LARGE $\br$  Form Factors}} \\
  \vspace{0.75in}
 Damir Becirevic \footnote{damir@qcd.th.u-psud.fr}
\vskip 0.5cm
{\small \it Laboratoire de Physique Th\'eorique et Hautes Energies\\
Universit\'e de Paris XI, B\^atiment 211, 91405 Orsay Cedex, France}\\
\vskip 0.8in
  ABSTRACT
 \vspace{0.8cm}
  \end{center}
\begin{quotation}
  \noindent
The bounds on the form factors for $\br$ decay are studied.
Constrained by lattice data and a constrained conformal mapping, the more
informations can be obtained for $A_1(q^2)$ form-factor which dominates the
decay rate at large $q^2$. Specifically, we confirm a moderately increasing
behavior of this form factor.
\end{quotation}

\end{titlepage}
\newpage
\section{Introduction}
%\vspace*{-0.35cm}

In the Standard Model, the couplings of quarks and $W$-boson are given in the
form of Cabibbo-Kobayashi-Maskawa ($CKM$) matrix elements. In this picture,
quark mixing and $CP$ - violation are closely related to each other. In order
to check whether this picture really describes $CP$ - violation and, more
generally, whether the Standard Model gives a proper description of weak decays
of hadrons, one must determine precise values of the $CKM$-matrix elements -
$\vert V_{ij} \vert$. One of the most poorly known among them is $\vert
V_{ub}\vert$. Naturally it is to be extracted from experimental data for the
semileptonic $b \to u$ transition. Experimentally, these processes are very
difficult to measure, because of the dominant $b \to c$ transition. Thus, $b
\to
u$ are clearly observable in a small fraction of the phase space, $i.e.$ beyond
the lepton momentum spectrum for $b \to c$. Theoretical description of the
inclusive $B \to X_{u} \ell \bar \nu_{\ell}$ decay is tremendously difficult
and suffers from large uncertainties. On the other side, there is a general
hope that $\vert V_{ub}\vert$ can be extracted from the exclusive channels,
particularly from $\br$. For reliable extraction of $\vert V_{ub}\vert$, we
must have precise experimental data on branching ratios as well as a hadronic
matrix element accurately calculated from $QCD$. Recently, $CLEO$ collaboration
\cite{cleo1,cleo2} observed and measured the branching ratios for this decay.
The reported values are:
\begin{equation}
 Br(\bar B^{0} \rightarrow \rho^{+} \ell^{-} \bar \nu_{\ell}) = \left\{
\begin{array}{c c}
(3.88 \pm 0.54 \pm 1.01) 10^{-4} & WSB \\
(2.28 \pm 0.36 \pm 0.59) 10^{-4} & ISGW
\end{array} \right .
\end{equation}
This branching ratio depend upon theoretical input for the efficiency
calculation. They used the quark model predictions ($WSB$ \cite{wbs} and $ISGW$
\cite{isgw}). We see that differences between two results are due to the models
they used, $i.e.$ theoretical uncertainties are important source of error.
Theoretically, the problem is how to calculate the hadronic matrix element and
the
corresponding weak transition amplitude since it receives hard non-perturbative
(low-energy) $QCD$ contributions. Unfortunately, up to now there is no
theoretical tool by which one can calculate the corresponding form factors
exactly in the whole physical region for this decay.

Most of the problems with quark model calculations are related to a lack of a
fully relativistic treatement of quark spins. In fact, they calculate a
specific form factor at one physical point and then assume the functional
dependence on $q^2$ in the whole physical region. The most popular assumption
is the polelike one. A step forward in calculating form factors was the chiral
perturbation theory of heavy hadrons, where the chiral and heavy quark
symmetries were used to construct a phenomenological lagrangian
\cite{gatto,nardulli}. However, only the small range of $q^2$ close to
$q^{2}_{max}$ can be covered, and again the ansatz on functional dependence of
form factors must be adopted. Different versions of the $QCD$ sum rules were
employed as well. These calculations allow to conclude about the functional
dependence, except when close to $q^{2}_{max}$. Still, the various $QSR$ give
different results. Finally, there are lattice $QCD$ simulations which, along
with the $QCD$ sum rules, are the only methods to treat non-perturbative $QCD$
in a consistent way (the nice review of the heavy mesons phenomenology from
lattice
$QCD$ can be found in Ref.\cite{martinelli}) . But, due to the $UV$-cutoff, the
relevant matrix elements are calculated for $m_{Q} \sim m_{c}$ and $q^2
\lesssim m_{c}^{2}$. $HQET$ is then used to control the extrapolation to
$m_{b}$ and consequently the values of accessible $q^2$ are restricted to a
small range close to $q^{2}_{max}$. However, this is the phase-space region
which is expected to contribute to a substantial fraction of the decay events
and is above the endpoint for charm production in the decay. For branching
ratio, we need the values of the hadronic matrix element at small $q^2$ too.
Then the corresponding form factors must be extrapolated to this region. To
perform this, we need an extrapolation law which is always the hypothesis we
make. The form factors values after
extrapolation to small $q^2$ are extrapolation-law dependent. In this respect,
the situation is not very different from the quark-model approach. Still by
lattice results we may constrain the unitarity bounds which is the subject of
this letter, and hence try to reduce the possible choice on the scaling laws
for extrapolation to small $q^2$.

The authors of
Ref.\cite{wise} proposed another model-independent way to extrapolate $\vert
V_{ub}\vert$ using the double Grinstein type ratio and more specifically the
following decay modes: $\br$, $B\to K^{*}\nu \bar \nu$, $D\to \rho \bar \ell
\nu_{\ell}$ and $D\to K^{*}\bar \ell \nu_{\ell}$. However, the rare decays
$B\to K^{*}\nu \bar \nu$ and $D\to K^{*}\bar \ell \nu_{\ell}$ have not been
observed and there is much to be learned on form factors.

In Sec.2 of this paper, we give the necessary definitions and motivations for
this analysis. In Sec.3 we generate the unitarity bounds on the form factors
which are constrained by lattice results in Sec.4. In Sec.5 we examine the
functional dependence $A_1(q^2)$ by additionaly constrained bounds. Concluding
remarks are given in Sec.5 and 6.
\section{Form factors}

The hadronic matrix element for the $\br$ decay is parametrised as:
\begin{eqnarray}
<\rho (p')|J^\mu |B(p)> &=& \frac{2 V(q^2)}{M + m} \epsilon ^{\mu \nu \alpha
\beta} p_{\nu}p'_{\alpha}\epsilon_{r \beta} - i (M +
m)A_{1}(q^2)\epsilon_{r}^{* \mu} \cr
&+& i \frac{A_{2}(q^2)}{M + m}(p + p')^{\mu}(\epsilon_{r}^{*}\ p) - i \frac{2
m}{q^2} (\epsilon_{r}^{ *} p) A(q^2)(p-p')^{\mu}
\end{eqnarray}
\noindent
where $J_\mu = (V_\mu - A_\mu) = \bar u \gamma_\mu (1 - \gamma^5) b$; $M$ and
$m$ are the masses of the $B$ and $\rho$ mesons, $p$ and $p'$ are their momenta
respectivelly; $q = p - p'$ \ is  the momentum transferred to the leptons and
$\epsilon^{r}$ is the polarization vector of the $\rho$-meson. The form factor
$A(q^2)$ does not contribute to the decay rate in the limit of massless
leptons. In this case ($i.e.\ \ell =e, \mu$ is a very good approximation) , we
have:
\begin{eqnarray}
\frac{d\Gamma}{dq^{2}}(\bar B^{0} \rightarrow \rho^{+} \ell^{-} \bar \nu_{
\ell}) = \frac{G^{2}|V_{ub}|^{2}}{192\pi^{3}M^{3}} q^2 \lambda^{1/2}(q^2)
\left[ \vert H_{+}(q^2)\vert ^{2} +
\vert H_{-}(q^2)\vert ^{2} + \vert H_{0}(q^2)\vert ^{2} \right].
\end{eqnarray}
$\lambda (t) = (t + M^{2} - m^{2})^{2} - 4 M^{2} m^{2}$ is the usual triangular
function.
$H_{0}$ and $H_{\pm}$ are the helicity amplitudes which come from the
longitudinally and transversely polarized $\rho$- mesons and are given by:
\begin{eqnarray}
H_{0}(q^2)&=& \frac{M+m}{2m\sqrt{q^2}} \left[ (M^2 - m^2 - q^2)A_1(q^2) -
\frac{\lambda (q^2)}{(M + m)^2}A_{2}(q^2)\right] \cr
H_{\pm}(q^2)&=& (M+m) A_1(q^2) \mp \frac{\lambda^{\frac{1}{2}}(q^2)}{M+m}
V(q^2)
\end{eqnarray}
The dynamics of this decay is described by $A_1(q^2)$, $A_2(q^2)$ and $V(q^2)$
- Lorentz invariant form factors, which are obviously functions of $q^2$
($\equiv t$). The physical region $0 \leq q^2 \leq q^{2}_{max}\ (20.3\ GeV^2)$
is very large.
In dispersion relations approach, the form factor $A_{1,2}(t)$ can be
associated to $J^P = 1^+$, while $V(t)$ to $J^P = 1^-$ intermediate state.
This suggests a nearest pole dominance assumption on the behaviour of the
form factors.
$ELC$\cite{elc} and $APE$\cite{ape} used this approach to extrapolate to $q^2
=0$ for all
form factors:
\begin{eqnarray}
f_{i}(q^2)=\frac{f_{i}(0)}{1-\frac{q^2}{M^{2}_{pole}}}
\end{eqnarray}
where $f_i = V, A_1, A_2$.

In Ref.\cite{narison1,narison2}, it was found that $A_1(t)$ decreases with t,
while $A_2(t)$ moderately increases. Ref.\cite{ball1} suggested that $V(t)$ is
consistent with the pole dominance, $A_1(t)$ decreases and $A_2(t)$ can be
fitted with the pole behavior. Light cone sum rules were used in
Ref.\cite{braun} and \cite{ball2} and they conclude that $V(t)$ is steeper than
pole, $A_1(t)$ increases but is flatter than pole and $A_2(t)$ is compatible
with pole behavior. $UKQCD$ \cite{ukqcd} in their analysis show that the pole
behavior for $A_1(t)$ is preferred (but with $m_{pole}=7^{+2}_{-1}\ GeV$).
Casalbuoni $et\ al.$ \cite{nardulli} keep the nearest-pole dominance for $V(t)$
and $A_2(t)$, while $A_1(t)$ is consisted of polar and polynomial terms, so
that their effective lagrangian approach leads to the so-called 'soft scaling':
\begin{eqnarray}
V,A_{2}(q^2_{max})\sim \frac{M_{H}+M_{V}}{\sqrt{M_{H}}} \ ;\
A_{1}(q^{2}_{max})\sim \frac{\sqrt{M_{H}}}{M_{H}+M_{V}}.
\end{eqnarray}
The common choice is the so-called 'hard scaling' which comes from $HQET$ at
leading order $1/M_{H}$ and in infinite heavy quark mass limit:
\begin{eqnarray}
V,A_{2}(q^2_{max})\sim \sqrt{M_{H}} \ ;\ A_{1}(q^{2}_{max})\sim
\frac{1}{\sqrt{M_{H}}}
\end{eqnarray}
These scaling laws (up to ${\cal O}(M_{H}^{-2})$ and log corrections) are used
by lattice groups for the extrapolation from $m_{Q} \sim m_{c}$ to $m_{b}$. A
very nice discussion about the scaling laws can be found in
Refs.\cite{nardulli,aoojc}.
The other way is to extrapolate first to $q^2=0$ and then to heavy masses. The
advantage is evident since the physical region is smaller and the results after
extrapolation to $q^2=0$ are not that much affected by the ansatz assumed for
the form factors' behaviour. But the $HQET$ scaling laws are valid only for the
$q^2 \simeq q^{2}_{max}$. The way out was recently pointed out in
ref.\cite{ball2} which states that one can extract from $QCD$ that all form
factors decrease at $q^2=0$ with heavy mass as $\sim M_{H}^{-3/2}$. We hope
that this would help us to reduce the errors in future lattice analyses.
There were also several quark models employed for a prediction of $\br$
form factors. For instance, in the framework of the light-cone formalism, a
relativistic treatement of spin was proposed in Ref.\cite{jauss}, but the form
factor values were accessible for $q^2 \leq 0$. Dispersion formulation of this
quark model was used to relate form factors from $q^2 \leq 0$ region to the
physical region by performing the analytic continuation \cite{melikhov}. A
relativistic quark model based on the quasipotential approach was discussed in
Ref.\cite{faustov}.
{}From this small list of results, in spite of the evident progress of $QSR$
and lattice results, we see that no definite conclusion on the functional
dependence of the form factors can be drawn. Let us derive unitarity bounds
for $V(t)$ and $A_{1}(t)$. In this letter, we concentrate to these two
form factors and more particularly to $A_{1}(t)$ which dominates the decay rate
at
large $t$. This sort of analysis was started by authors of Ref.\cite{okubo},
and applied to $K \to \pi \ell \nu_{\ell}$ decay in Ref.\cite{bourely}. In
$B$-physics, Refs.\cite{grinstein,rafael,caprini,neubert,all} used it for the
heavy-to-heavy meson semileptonic decays.
The idea to employ the method for heavy-to-light decays was first discussed in
Ref.\cite{boyd}. So, most of the material discussed in this paper can be found
in the above references.

\section{Unitarity bounds}

The starting point is the two-point function:
\begin{eqnarray}
\Pi _{\mu \nu}^{V,A} \equiv i\int d^{4}x e^{iqx} <0|T(J^{V,A}_{\mu} (x)J^{V,A
\dagger}_{\nu
}(0))|0> =
(q^{\mu}q^{\nu} - q^{2}g^{\mu \nu})\Pi^{V,A}_{T}(q^{2}) + g^{\mu
\nu}\Pi^{V,A}_{L}(q^{2})
\end{eqnarray}
In $QCD$, both sides of this equation satisfy once subtracted dispersion
relations:

\begin{equation}
\chi^{T,L}_{V,A}(Q^{2}) = \frac{\partial \Pi_{T,L}^{V,A}(q^{2})}{\partial
q^{2}}
\arrowvert_{q^{2}=-Q^{2}} = \frac{1}{\pi} \int_{0}^{\infty} \frac{Im
\Pi_{T,L}^{V,A}(t)}{(t + Q^{2})^{2}}dt
\end{equation}

For $Q^2=0$, we are far from the region where the currents can create
resonances ($(m_{Q} + m_{q})\Lambda_{QCD} << (m_{Q} + m_{q})^{2} + Q^{2}$), so
that $\chi_{T,L}^{V,A}$ can be reliably calculated by means of perturbative
$QCD$.
The spectral functions $Im \Pi_{T,L}^{V,A}$ can be obtained from the unitarity
relation:
\begin{eqnarray}
(q^\mu q^\nu - q^2 g^{\mu \nu})Im \Pi^{T}_{V,A}(t + i\epsilon)\ +\  g^{\mu
\nu}Im \Pi^{L}_{V,A}(t + i\epsilon) =
\nonumber
\end{eqnarray}
\begin{eqnarray}
\frac{1}{2} \sum_{\Gamma} \int d\rho_{\Gamma} (2\pi)^{4}\delta
(q-p_{\Gamma})<0\vert J_{V,A}^{\mu}(0)\vert \Gamma><\Gamma \vert
J_{V,A}^{\nu}(0)\vert 0>
\end{eqnarray}
\noindent
where $\Gamma$ are all possible hadron states with appropriate quantum numbers,
and the integration goes over the phase space allowed to each intermediate
state. We proceed as in Ref.\cite{grinstein}. For $\mu = \nu$, this is the sum
of positive definite terms. By
concentrating on the $B\rho$-intermediate state, we obtain the strict
inequalities. From crossing symmetry, we know that the $B\rho \to vacuum$
matrix element is described by the same set of the form factors as in (2), but
in the different region: ($(M+m)^2 \leq q^2 \leq \infty$), $i.e.$ on the cut.
Taking the space-space components of the
unitarity relation, we obtain the same combinations of the form factors as
those entering the decay rate expression. So, it suffices to take:
\begin{equation}
\chi_{V,A} = \left( \chi_{V,A}^{T}(Q^2) + \frac{1}{2} \frac{\partial}{\partial
Q^2} \chi_{V,A}^{L}(Q^2)\right)_{Q^2=0} .
\end{equation}
Finally, we have,
\begin{eqnarray}
Im \Pi^{ii}_{V}(t) \geq \frac{2}{3 \pi (M+m)^2}
\frac{\lambda^{\frac{3}{2}}(t)}{t} \vert V(t)\vert^2 \theta(t-t_{+})\\
\nonumber \\
Im \Pi^{ii}_{A}(t) \geq \frac{1}{12 \pi t}\lambda^{\frac{1}{2}}(t) \left[
2(M+m)^2 \vert A_1(t)\vert^{2} + \vert H_{0}(t)\vert^2 \right] \theta(t-t_{+})
\end{eqnarray}
\noindent
where $t_{\pm} = (M \pm m)^2$. Inserting these functions in
the dispersion relation at $Q^2 = 0$:
\begin{eqnarray}
\chi_{V,A} = \frac{1}{\pi} \int_{0}^{\infty} dt \frac{Im
\Pi^{ii}_{V,A}(t)}{t^3}
\end{eqnarray}
\noindent
we obtain the set of inequalities:
\begin{eqnarray}
\frac{1}{12 \pi \chi_{V} t_{+}} \int_{t_{+}}^{\infty} dt
\frac{\lambda^{\frac{3}{2}}(t)}{t^4} \vert V(t)\vert^2 \leq 1 \\ \nonumber
\\
\frac{1}{12 \pi \chi_{A}} \int_{t_{+}}^{\infty} dt
\frac{\lambda^{\frac{1}{2}}(t)}{t^4} \left\{ 2 t_{+} \vert A_{1} \vert^{2} +
\vert H_{0} \vert^2 \right\} \leq 1
\end{eqnarray}
\noindent
For bounds on the form factors, we satisfy ourselves by calculating
$\chi_{V,A}$ at leading order:

\begin{equation}
\chi_{V,A} = \frac{N_{c}}{4(2\pi)^{D/2}} \Gamma (3-\frac{D}{2}) \int_{0}^{1} dx
\frac{x^2(1-x)^2[(D+2)(m_Q\mp m_q)(m_Q x \mp m_q(1-x)) + 8 m_q
m_Q]}{(m_{Q}^{2}x + m_{q}^{2}(1-x))^{4-\frac{D}{2}}}
\end{equation}

\vspace{3mm}
\noindent
which in our case ($m_{u}=0$) gives: $\chi_{V} = \chi_{A} = \frac{3}{4 \pi
m_{b}^{2}}$.

Actually, we have three inequalities which constrain form factors in the
unphysical kinematic region. To obtain the bounds on the physically interesting
form factors for $\br$, we perform the conformal mapping:

\begin{eqnarray}
\frac{1+z}{1-z} = \sqrt{\frac{t_{+}-t}{t_{+}-t_{-}}}
\end{eqnarray}
by which the whole complex $t$-plane is mapped onto the unit disc $\vert z
\vert \leq 1$. More specifically: $t_{-} \leq t \leq t_{+}$ is mapped into the
segment of the real axis $-1 < z \leq 0$, while $0 \leq t \leq t_{-}$ is mapped
into $0 \leq z < 1$. In other words, the physical region for $B\rho \to vacuum$
transition lies on the unit circle, and for $\br$ decay on the right segment of
the real axis inside the unit circle.
In the $z$ - plane, the inequalities (16,17) become:
\begin{eqnarray}
\frac{1}{2\pi i} \int_{\cal C} \frac{d z}{z} \vert \Phi_{i}(z)f_{i}(z)\vert^{2}
\leq 1.
\end{eqnarray}
Here, we wrote generically $f_{i} = V, A_{1}$ and their corresponding functions
$\Phi_{i}$; $\cal{C}$ is the unit circle. The procedure for obtaining the
functions $\Phi_{i}(z)$ is well-known and is the solution of Dirichlet's
boundary problem (\cite{okubo,rafael}): its value is known on the circle $i.e.$
$\vert \Phi_{i}(e^{i\theta})\vert^2$. Solutions are:

\begin{eqnarray}
\Phi_{V}(z) = \sqrt{\frac{2}{3\pi \chi_{V}}} \frac{32M^2 m^2}{(M + m)^5}
\frac{(1+z)^2}{(1-z)^{\frac{9}{2}}}\left( 1 + \frac{2\sqrt{Mm}}{M+m}
\frac{1+z}{1-z} \right)^{-4} \\
\Phi_{A_{1}}(z) = \sqrt{\frac{1}{3\pi \chi_{V}}} \frac{8M m}{(M + m)^3}
\frac{1+z}{(1-z)^{\frac{5}{2}}}\left( 1 + \frac{2\sqrt{Mm}}{M+m}
\frac{1+z}{1-z} \right)^{-4} \\ \nonumber
\end{eqnarray}
\noindent
The functions $\Phi_{i}(z)$ are analytic everywhere inside the unit circle.
{}From the point of view of analyticity, the problems arise with the
form factors. All singularities situated above threshold $t_{+}$ can be
absorbed in the phase that can be added in redefinition of $\vert
\Phi_{i}(e^{i\theta})\vert$ and eventually will not contribute to our bounds.
But the singularities inside the gap $t_{-}\leq t\leq t_{+}$ are important.
First of all, there are two poles, one at $t=(5.32\  GeV)^2 \to
z_{pole1}=-0.1666$, contributing to $V(t)$ and the other at $t=(5.73\  GeV)^2
\to z_{pole2}=-0.3514$, contributing to $A_{1}(t)$. Since we do not know the
residua
of the form factors at these poles, we will simply remove them. This can be
achieved by introducing the Blaschke factors:
\begin{eqnarray}
P_{V}=\frac{z-z_{pole1}}{1-z z^{*}_{pole1}} \quad
P_{A_{1}}=\frac{z-z_{pole2}}{1-z z^{*}_{pole2}}
\end{eqnarray}
Since the Blaschke factors $P_{V}$ and $P_{A_{1}}$ are unimodular, after
inserting them into (20) ($\Phi_i(z) \to P_i(z)\Phi_i(z)$) the inequalities
remain the same. In fact, this biases our analysis: if we knew the values of
residua, the form factors would be precisely determined for the large values of
$q^2$, or at least our bounds would be very narrowed in the whole physical
region.

There is also a problem to incorporate subthreshold singularities. They are
expected in the analysis of the $A_{1}(t)$ form factor at $(M_{B^{**}}+n
m_{\pi})^2$. So, there are two branch points below the threshold ($z_{1}=
-0.4671 ,z_{2}= -0.7061$) . Their effect is negligible in our case, as it can
be verified by applying the models discussed in Refs.\cite{neubert} and
\cite{grinstein}.
It turns out that the bounds on $A_{1}(t)$ would be relaxed by no more than
$1\%$.
The last step in deriving the bounds is the construction of the inner product:

\begin{equation}
(g_{i},g_{j}) = \int_{\cal C} \frac{dz}{2\pi iz}  g_{i}^{*}(z) g_{j}(z).\\
\nonumber
\end{equation}
\noindent
and choose $g_{1}(z)=\Phi_{i}(z)P_{i}(z)f_{i}(z)$ and $g_{2}(z)=(1 -
zz_{1}^{*})^{-1}$. Then, from the positivity of the inner product, we obtain
that the determinant of the $(g_{i},g_{j})$ matrix is positive, $i.e.$:

\begin{equation}
\left| \begin{array}{cc}
1 & f_{i}^{*}(z_{1})\Phi_{i}^{*}(z_{1}) P_{i}^{*}(z_{1}) \\
f_{i}(z_{1})P_{i}(z_{1})\Phi_{i}(z_{1}) & \frac{1}{1 - |z_{1}|^{2}}
\end{array} \right|  \geq 0  , \quad \forall z_{1}\in {\mathrm Int} \cal C
\end{equation}
or
\begin{equation}
\vert V(z)\vert \leq \frac{1}{P_{V}(z)\Phi_{V}(z)} \frac{1}{1 - \vert
z\vert^{2}} , \quad \vert A_{1}(z)\vert \leq
\frac{1}{P_{A_{1}}(z)\Phi_{A_{1}}(z)} \frac{1}{1 - \vert z\vert^{2}}
\end{equation}

\vspace{5mm}
We see from the picture (Fig.1) that such bounds are not at all restrictive
($\vert V(t)\vert \leq 16, \vert A_1(t)\vert \leq 18.3$). Our nice exercise did
not lead to any reasonable restriction on the form factors. To constrain them
more we can use some of the form factor values that we have on our disposal.
Similar analysis was performed for the heavy-to-light transition but for the
case of
$B \rightarrow \pi \ell \bar\nu_{\ell}$ in Refs.\cite{lellouch,damir}.

$$ \epsfbox{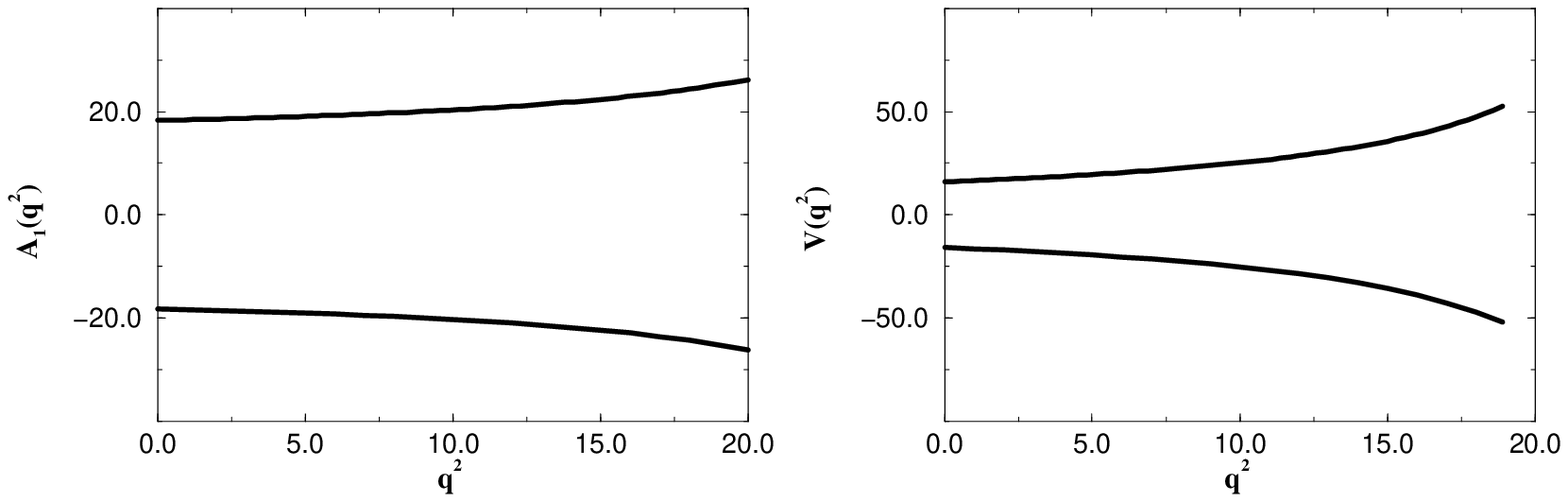}$$
\vspace{1cm}
\centerline{{\bf Fig.1}: \it Unitarity bounds on the form
factors $A_1(q^2)$ and $V(q^2)$.}
\\

\section{Lattice constrained bounds}

 To incorporate $n$ lattice results in our analysis, we define $g_{i}(z)=(1 -
zz_{i}^{*})^{-1}$\\ ($i=1,\ldots ,n$) and construct the matrix $(n+2) \times
(n+2)$ whose determinant is positive, $i.e.$:

\vskip .03cm
\begin{equation}
\left| \begin{array}{ccccc}
1 & f_{i}^{*}(z)\Phi_{i}^{*} (z) P_{i}^{*} (z)& f_{i}^{*}(z_{1})\Phi_{i}^{*}
(z_{1})P_{i}^{*} (z_{1})&  ... & f_{i}^{*}(z_{n})\Phi_{i}^{*} (z_{n}) P_{i}^{*}
(z_{n})\\

f_{i}(z)P_{i} (z) \Phi_{i} (z) & \frac{1}{1 - |z|^{2}} & \frac{1}{1 -
zz_{1}^{*}}&
... & \frac{1}{1 -  zz_{n}^{*}} \\
  &  &  &  &  \\
  & \cdots  &  \cdots & \cdots &  \\
f_{i}(z_{n})P_{i} (z_{n}) \Phi_{i} (z_{n}) &\frac{1}{1 - z_{n}z^{*}}&
\frac{1}{1 -
z_{n}z_{1}^{*}} & ... &  \frac{1}{1 - |z_{n}|^{2}} \end{array} \right|  \geq  0
\nonumber
\\
\end{equation}
\vskip .3cm

For our purpose, we take three (or two) lattice data of the form factors $i.e.$
the results for the $B$-meson at rest, which are more precise. Since the
$UKQCD$ data are very accurate, we take their results (all technical details
can be found in Ref.\cite{ukqcd}). In the table below, the lattice results of
three collaborations are given \footnote{The Wuppertal group \cite{wup} also
studied this decay by extrapolating first to $q^2=0$ and then to the heavy
mass.}:

\vspace{5mm}

\begin{tabular}{||c||c|c|c|c||}   \hline
 & $q^2$ $[GeV^2]$&$A_{1}(q^2)$&$A_{2}(q^2)$&$V(q^{2})$ \\ \cline{2-5}
\underline{UKQCD}\cite{ukqcd} & $20.3$ & $0.46^{+2}_{-3}$ & - & - \\
$24^3 \times 48$ lattice&
$17.5^{+2}_{-2}$&$0.43^{+2}_{-2}$&$0.8^{+2}_{-2}$&$1.6^{+1}_{-1}$ \\
$\beta = 6.2$ &
$15.3^{+3}_{-3}$&$0.39^{+3}_{-2}$&$0.7^{+2}_{-1}$&$1.2^{+1}_{-1}$ \\
Clover Action&$16.7^{+2}_{-2}$&$0.38^{+3}_{-3}$&$0.6^{+3}_{-3}$&$1.5^{+1}_{-1}$
\\
&$14.4^{+3}_{-3}$&$0.39^{+6}_{-5}$&$0.7^{+3}_{-2}$&$1.4^{+3}_{-2}$ \\  \hline
\hline
\underline{APE}\cite{ape}  & $q^2$ $[GeV^2]$&$A_1(q^2)$&$A_2(q^2)$&$V(q^2)$ \\
\cline{2-5}
$18^3 \times 64$ lattice& $20.3$& $0.43\pm .08$ &-&-    \\
$\beta = 6.0$ & $17.6$&$0.48\pm .16$&$0.51\pm 0.50$  &$1.6\pm .6$  \\
Clover Action&$16.6$&$0.70\pm .37$  &$0.55\pm 0.70 $ &$1.2\pm 2.2$  \\
 &$13.5$&$1.06\pm .98$  &$1.05\pm 1.10 $ &$1.2\pm 4.1$  \\  \hline \hline
\underline{ELC}\cite{elc}& $q^2$ $[GeV^2]$&$A_1(q^2)$&$A_2(q^2)$&$V(q^2)$ \\
\cline{2-5}
$24^3 \times 60$ lattice& $20.3$&$0.60(6)$  & - &  - \\
$\beta = 6.4 $& $15.8$&$0.53(9)$&$0.49(22)$&$0.77(16)$ \\
Wilson Action&$12.5$&$0.43(25)$&$0.29(37)$&$0.56(20)$ \\  \hline
\end{tabular}
\vspace{5mm}

\noindent
We see from the table, that the lattice results are quite far from being
satisfactory when a behavior of form factors is to be studied. Results quoted
above are concentrated in the vicinity of $q^{2}_{max}$, and must be
extrapolated to small $q^2$. With our constrained bounds, we want to
restrict the values of the form- factors in the region of small and
intermediate $q^2$. With lattice data incorporated, the constrained
bounds are obtained from (27) (thus, we take first
three/two results from the table above). Resulting bounds are shown on the
pictures (Fig.2).
\vspace{-4mm}
$$ \epsfbox{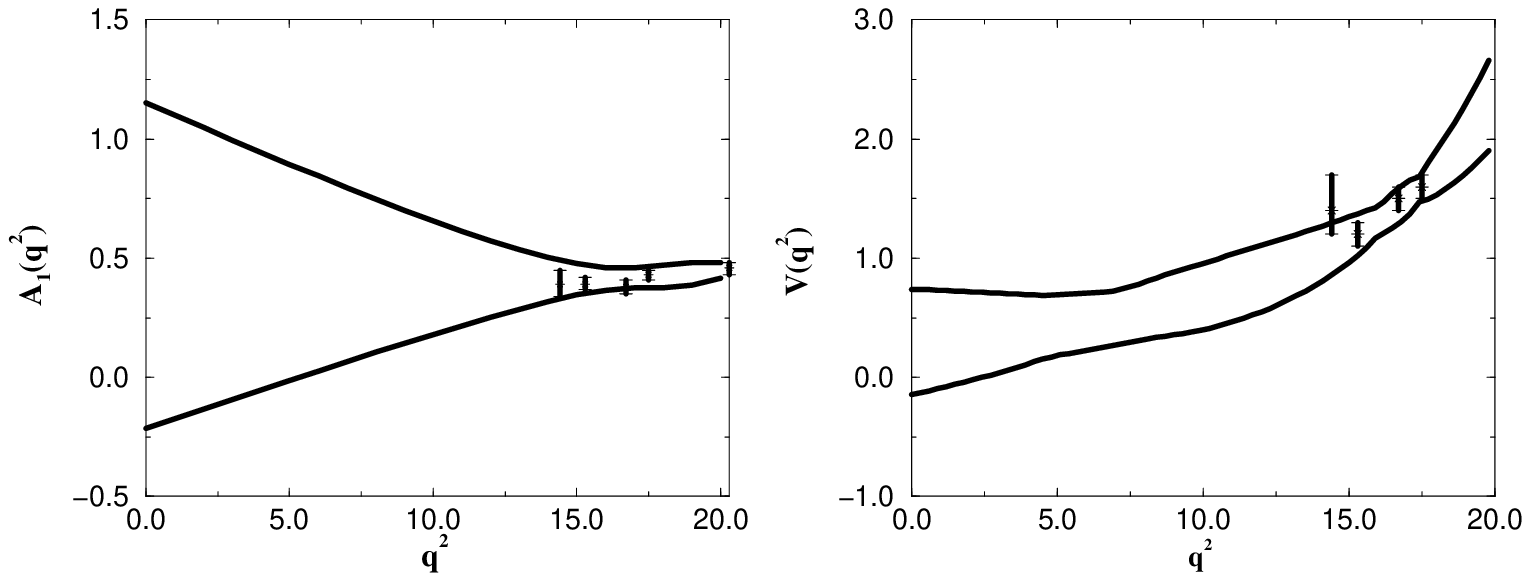}$$
\parbox{6.2in}{{\bf Fig.2}: \it Lattice constrained unitarity bounds on the
form
factors $A_1(q^2)$ and $V(q^2)$ plotted with lattice UKQCD results.}\\

\vspace{8mm}
{}From the pictures, we see immediately how stronger the bounds are (for
instance, $-0.22\leq A_1(0)\leq 1.14$ and $-0.17\leq V(0)\leq 0.72$). They are
far more restrictive than the previous ones but the problem of the form factors
behavior
remains unsolved. Beside statistical, we did not incorporate other errors in
this analysis\footnote{The complete treatement of errors is discussed in
Ref.\cite{lellouch}}. This will be done in the next section. If we try to see
typical ans\"{a}tze taken for extrapolations to $q^2=0$, we see that most of
them fall inside the region of allowed values for the form factors at hand.
Still, we can try to modify our analysis in the step (19).

\section{More constrained bounds}

Again, we perform the conformal mapping:

\begin{eqnarray}
\frac{1+z}{1-z} = \sqrt{\frac{t_{+}-t}{N (t_{+}-t_{-})}}
\end{eqnarray}

\noindent
where $N$ is a constant. The endpoints of the physical region are mapped as;
\begin{eqnarray}
t = 0  &\longmapsto&  z_{max}= \frac{\sqrt{t_{+}} -
\sqrt{N(t_{+}-t_{-})}}{\sqrt{t_{+}} + \sqrt{N(t_{+}-t_{-})}} \\
t = q^2_{max}  &\longmapsto&  z_{min}=-\left( \frac{\sqrt{N} - 1}{\sqrt{N} +
1}
\right)
\end{eqnarray}

\noindent
The inequalities that we derived in Sec.2  remain the same, but the functions
(21,22) now become:

\begin{eqnarray}
\Phi^{N}_{V}=\frac{16 M^{2} m^{2} N}{\sqrt{3\pi \chi_{V}}(M + m)^5} (1 +
z)^{\frac{7}{2}}(1 - z)^{-\frac{9}{2}} \left( 1 + \frac{1 - z}{\sqrt{N}(1 + z)}
\right)^{\frac{3}{2}} \left( 1 + \frac{2\sqrt{NMm}}{M + m} \frac{1 + z}{1 -
z}\right)^{-4} \\
\Phi^{N}_{A_1}=\frac{4\sqrt{2} M m N}{\sqrt{3\pi \chi_{A}}(M + m)^3} (1 +
z)^{\frac{3}{2}}(1 - z)^{-\frac{5}{2}} \left( 1 + \frac{1 - z}{\sqrt{N}(1 + z)}
\right)^{\frac{1}{2}} \left( 1 + \frac{2\sqrt{NMm}}{M + m} \frac{1 + z}{1 -
z}\right)^{-4}
\end{eqnarray}
\vspace{4mm}
\noindent
Of course, for $N=1$ we recover (21,22).

Let us choose $N$ in such a way that $ z_{max} = -z_{min}$. This gives $N
\simeq 1.5$ and consequently $z_{max} = -z_{min} =
0.1011$. This means that $z$ is a small kinematic parameter in the whole
physical region, and that we can Taylor expand the functions
$\Phi_{i}(z)P_{i}(z)f_{i}(z)$, around $z = 0$ $i.e.$:
\begin{equation}
f_{i}(z) = \frac{1}{P_{i}(z)\Phi_{i}(z)} {\sum_{n=0}^{\infty}{a_{n}z^{n}}}
\qquad
\end{equation}
and from inequalities (17) we extract the additional constraint:
\begin{equation}
\sum_{n=0}^{\infty} |a_{n}|^{2} \leq 1
\end{equation}
Actually, the coefficients in the series $a_{i}$ can only be obtained from the
data (In the case of $B \to D^{(*)} \ell \bar\nu_{\ell}$ one coefficient is
obtained by the help of $HQS$ which gives the absolute normalization of the
form factors at $q^{2}_{max}$. In our case, we do not have such an advantage.).
It implies that we have to truncate our series. By taking the first $k$ terms,
and using the Schwartz inequality and condition (34), we can estimate the
truncation error as:

\begin{eqnarray}
\Delta_{tr}[f_{i}] = \max{|f_{i}(z) - f_{i}^{k}(z)|} &\leq& \max
{\frac{1}{\vert P_{i}(z)\Phi_{i}(z)\vert}\sqrt{\sum_{n=k+1}^{\infty} \vert
a_{n}\vert^{2}z^{2n}}} \cr
&<& \max {\frac{1}{\vert P_{i}(z)\Phi_{i}(z)\vert}\frac{z^{k+1}}{\sqrt{1-z^2}}}
\end{eqnarray}

For the truncation errors in our case, we have:

\vskip 3mm
\begin{center}
\begin{tabular}{|c|c|c|c|} \hline
k& $\Delta_{tr}[V(t)]$ &$\Delta_{tr}[A_{1}(t)]$  &$\Delta_{tr}[H_{0}(t)]$    \\
\hline
0 & 1.761 & 2.672 & 32.226 \\ \hline
1 & 0.178 & 0.269 & 3.267 \\ \hline
2 & 0.018 & 0.027 & 0.330 \\ \hline
\end{tabular}
\end{center}
\vskip 3mm

As it was already mentioned, for conservative bounds we take three/two lattice
data for the form factors $A_1(t)\ (k=2)$ / $V(t)\ (k=1)$. We display the
resulting plot for $V(t)$ even though the truncation error is 'large'. In
Ref.\cite{grinstein}, it was noticed that we can incorporate other
uncertainties by choosing the constant on the r.h.s. of (34) greater than one.
Namely, in calculation of $\chi_{V,A}(0)$, we may include $\cal{O}(\alpha_{s})$
corrections which contribute $\sim 18\%$ of the one loop contribution. This
corrections, and uncertainties on $m_{b}=4.8^{+2}_{-2}\ GeV$, as well as
contributions of the subthreshold singularities relax the constraint (34) by no
more than $20\%$, $i.e.$ $\sum_{n=0}^{\infty} |a_{n}|^{2} \leq 1.2$. By this
condition, we can determine the value of the coefficient $a_3$($a_2$) in the
expansion for $A_1(z)$($V(z)$). Since we consider the bounds, we take the sign
which leads to more conservative bounds. With this in mind, we obtain:

\vskip 3mm
\begin{center}
\begin{tabular}{|c||c|c|c|c| c c|c||c|c|c|} \cline{1-5} \cline{8-11}
$A_1(z)$& $a_0$ &$a_1$  &$a_2$ &$a_3$ & & &$V(z)$ & $a_0$ &$a_1$  &$a_2$  \\
\cline{1-5} \cline{8-11}
$upper$ & 0.0178 & -0.0393 & -0.3474&-1.0379& & &$upper$&0.0155&-0.0412&-1.0946
\\ \cline{1-5} \cline{8-11}
$lower$ &0.0141 & -0.0956& -0.7279&0.8129& & &$lower$&0.0123&-0.0589&1.0938 \\
\cline{1-5} \cline{8-11}
\end{tabular}
\end{center}
\vskip 3mm

Besides statistical errors, in Ref.\cite{ukqcd} the systematic errors were
estimated. Apart from quenching, they quote $11\%$ for $A_1(t)$ and $15\%$ for
$V(t)$ of systematic errors. We relax the resulting bounds by the value of
these errors. Final bounds are plotted on Fig.3. \footnote{We used also $ELC$
and $APE$ results to generate the bounds. Naturally, these bounds are much
weaker.}
$$ \epsfbox{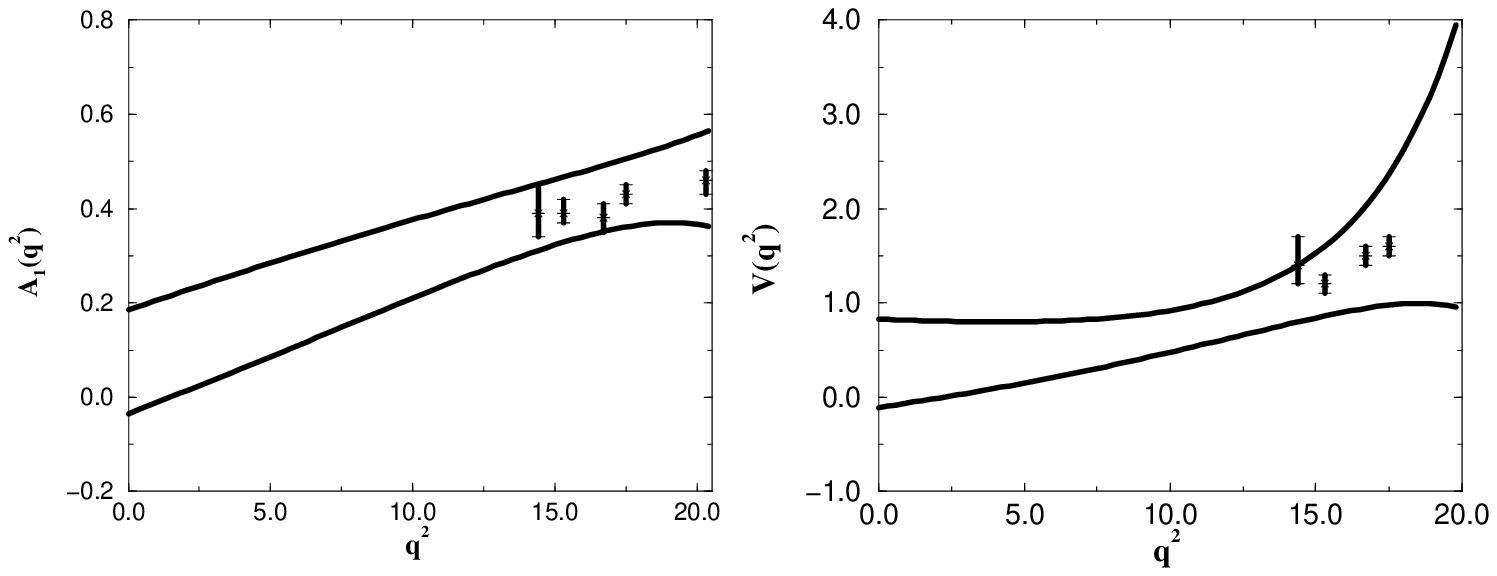}$$
\parbox{6.2in}{{\bf Fig.3}: \it More constrained bounds on the form factors
$A_1(q^2)$ and $V(q^2)$ (see text) and lattice UKQCD results (bounds are
relaxed by the value of systematic errors).}
\\

\vspace{8mm}
{}From these figures, we may see first that $A_1(t)$ increases with $t$. We
also notice that $A_1(0)$ is small and $A_1(0)\leq 0.18$. On Fig.4, we plot
some predictions on functional dependence $A_1(t)$ against our bounds: On
Fig.4c, the comparison with the nearest pole behavior \cite{wbs} shows that
$A_1$ must be flatter. Lattice groups ($UKQCD$, $APE$ and $ELC$) tried to fit
with the pole behavior but with the masses $m_{pole}= 7^{+2}_{-1}\ GeV;
5.99\pm0.62 GeV; 5.62\pm 0.11 GeV$ respectivelly. If we want $A_1(t)$ to fall
within our bounds, the pole mass should be $m_{pole}\geq 6.5\ GeV$ which is
quite bigger than the nearest pole mass. This can be interpreted as if radial
and orbital excitations had more impact on the form factor behavior in the
physical region. For $B \to \pi \ell \bar \nu_{\ell}$ in Ref.\cite{burdman},
the authors tried to study such effects by taking into account the first few
orbital
excitations (uncertainties are large since the values of $f_{B_i}$ are almost
unknown). On Fig.4b, we plot $UKQCD$ prediction by taking $m_{pole}= 7\ GeV$.
In Ref.\cite{nardulli}, they obtain that $A_1(t)$ should be the nearest pole
plus a polynomial term and more specifically, they take a polynomial term to be
a constant. This is plotted on Fig.4a and it seems that linear term should be
included too. Finally, on Fig.4d the functional dependence $A_1(t)$ predicted
by light cone sum rules is plotted. \parbox{6.2in}
{$$ \epsfbox{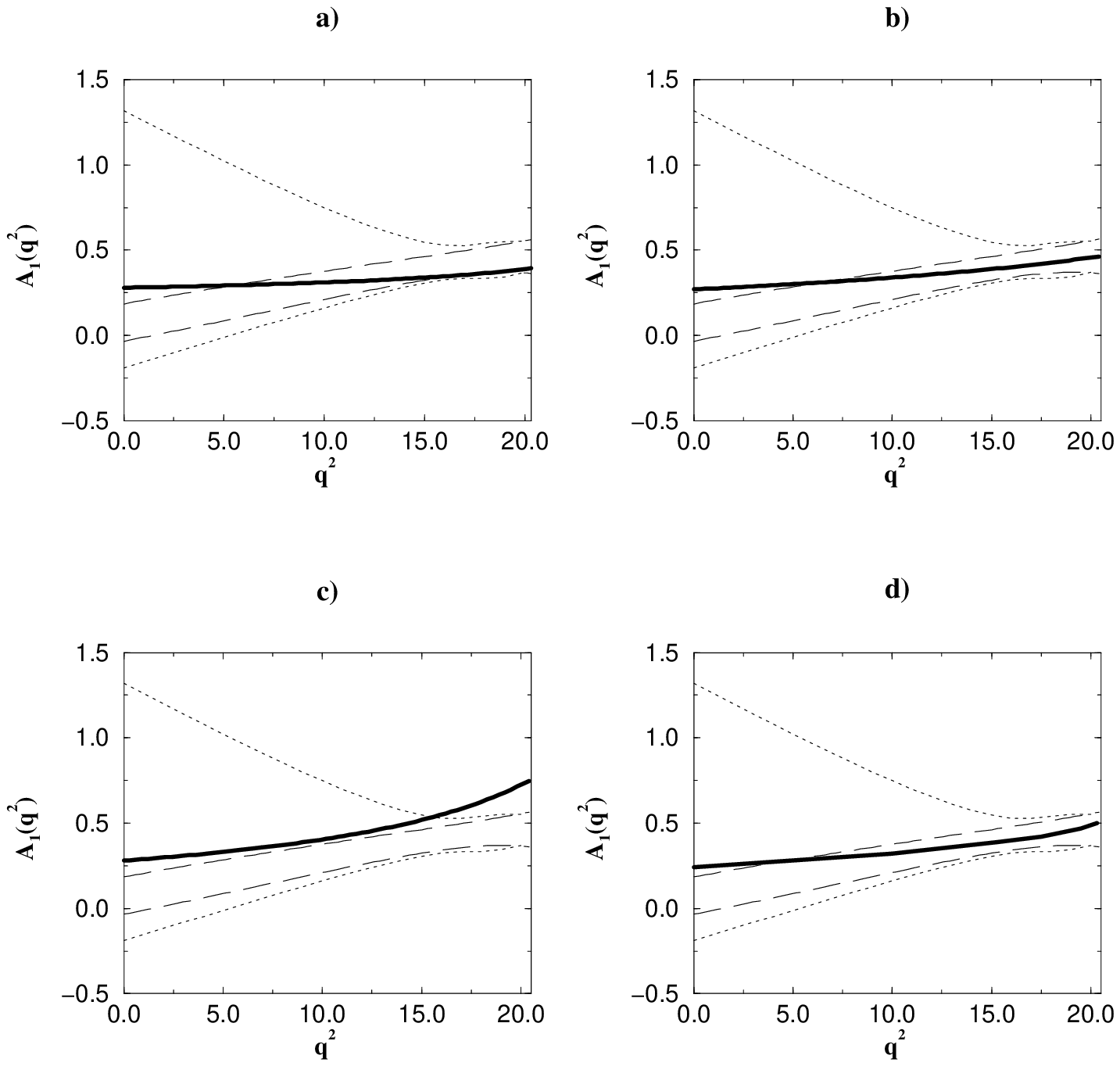}$$ \\
{\bf Fig.4}: \it a)Effective lagrangian (CHPTHH, Ref.\cite{nardulli})
prediction of $A_1(q^2)$; b) UKQCD pole fit (see text and Ref.\cite{ukqcd}); c)
Nearest pole dominance \cite{wbs}; d) Light-cone QCD sum rules prediction of
$A_1(q^2)$; are plotted against the bounds from Sec.4 (dotted) and Sec.5
(dashed)}
\vspace{1cm}

\noindent
Note that the bounds generated in Sec.4 are relaxed by the value of the
systematic errors.

\section{Conclusion}
In this paper, we studied the unitarity bounds on the form factors for $\br$
decay. Form factors and perturbative calculation of a two-point function are
related to each other by crossing symmetry and a dispersion relation. By
conformal mapping we obtain the 'unrestrictive' bounds on the physically
interesting form factors. The presence of poles was taken into account by
corresponding Blaschke factors which simply remove them, since we do not know
the residua $i.e.$ $g_{\rho BB_{pole}}f_{pole}$. The bounds obtained in this
way are
constrained by the lattice results (more specifically, results obtained by
$UKQCD$). We took only these values where the heavy meson was at rest, and with
errors included to make our bounds more conservative. Besides the bounds
obtained in this way, we wanted to study the functional dependence of the form
factors. With more constrained analysis, we see that with present data we can
study only $A_1(t)$ which dominates the decay rate for large values of $t$. We
confirm that $A_1(t)$ cannot decrease, but moderately increases, $i.e.$ it is
flatter than the nearest pole dominance hypothesis would give. The value of
$A_1(0)$ bounded in this way, is surprisingly small ($A_1(0)\leq 0.18$). In
these final bounds, we did not include the errors of quenching. If we take them
into account and relaxe our bounds by conservative $10 \%$, we would have
$A_1(0)\lesssim 0.20$. For $V(t)$ we can not conclude whether it behaves like
pole or it is steeper due to the truncation error that we have. In fact both,
pole and double-pole functional dependence, are fully inside the region allowed
by our bounds.  The analysis performed in this paper is even more needed for
$B\to K^{*}\gamma$
form factors. This work is in progress.

\vskip 1.5cm
{\bf Acknowledgement}
\vskip 1.0cm

It is a pleasure to thank A. Le Yaouanc and J.P. Leroy for their advices and
motivating discussions.

\end{document}